\newcommand{\be}{\begin{equation}}
\newcommand{\ee}{\end{equation}}
\newcommand{\bea}{\begin{eqnarray}}
\newcommand{\eea}{\end{eqnarray}}

\documentstyle[12pt,epsfig]{article}

\begin{document}
\renewcommand{\thefootnote}{\fnsymbol{footnote}}
\thispagestyle{empty}
%\rightline{LNF-97/016 (P)}
%\rightline{gr-qc/9704039 \vspace{1cm}}
\begin{titlepage}
\begin{center}
%%%%%   {\Large\bf EFFECTS OF THE NUMBER OF SUBBANDS ON THE RG
%%%%%   SOLUTIONS AND THE PROPERTIES OF MULTI WALLED NANOTUBES}

{\Large\bf LARGE $N$ EFFECTS AND RENORMALIZATION OF THE
LONG-RANGE COULOMB INTERACTION IN CARBON NANOTUBES}
\end{center}
\vskip 0.6truecm
\begin{center}
{\large\bf S. Bellucci}\footnote{bellucci@lnf.infn.it } \vspace{0.5cm} \\
{\it INFN-Laboratori Nazionali di Frascati, \\
 C.P. 13, 00044 Frascati, Italy \vspace{0.5cm} }\\
{\large\bf J. Gonz\'alez\footnote{imtjg64@pinar2.csic.es }}
  \vspace{0.5cm} \\
{\it Instituto de Estructura de la Materia,\\
         Consejo Superior de Investigaciones Cient{\'\i}ficas,\\
Serrano 123, 28006 Madrid, Spain} \vspace{0.5cm} \\
{\large\bf P. Onorato\footnote{pasquale.onorato@libero.it }}
  \vspace{0.5cm} \\
{\it INFN-Laboratori Nazionali di Frascati, \\
 C.P. 13, 00044 Frascati, Italy} \\
{\it and} \\
{\it Dipartimento di Scienze Fisiche, \\
 Universit\`{a} degli
Studi di Napoli ``Federico II'',\\
Via Cintia, I-80126 Napoli, Italy} \vspace{0.5cm}
\end{center}
\vskip 0.6truecm  \nopagebreak
\begin{abstract}
\noindent
We develop a dimensional regularization approach to deal with
the low-energy effects of the long-range Coulomb interaction in
1D electron systems. The method allows us to avoid
the infrared singularities arising from the long-range Coulomb interaction
at $D = 1$, providing at the same time insight about the
fixed-points of the theory. We show that the effect of
increasing the number $N$ of subbands at the Fermi level is
opposite to that of approaching the bare Coulomb interaction in
the limit $D \rightarrow 1$. Then, we devise
a double scaling limit, in which the large $N$ effects
are able to tame the singularities due to the long-range
interaction. Thus, regular expressions can be obtained for all
observables right at $D = 1$, bearing also a dependence on the
doping level of the system.
Our results imply a variation with $N$ in the value of the exponent
for the tunneling density of states, which is in fair agreement
with that observed in different transport experiments involving
carbon nanotubes.
As the doping level is increased in nanotubes of large radius
and multi-walled nanotubes, we predict
a significant reduction of order $N^{-1/2}$ in the
critical exponent of the tunneling density of states.

\end{abstract}
\newpage
\end{titlepage}
\renewcommand{\thefootnote}{\arabic{footnote}}
\setcounter{footnote}0
\section{Introduction}

In the last decade carbon-based devices have been largely
studied as a possible new frontier of microelectronics. The limits
of further miniaturization (predicted by Moore's law) have
increased the research toward the development of molecular
electronics and new efforts of scientists have been stimulated by
the progress in carbon technology, in order to build a carbon
based microelectronics \cite{lbac,cbmac}. Carbon nanotubes are
the bricks of this new building and recent experiments have
revealed that they are also excellent systems for the
investigations of electronic transport in one dimension.
It has been shown that the single-walled nanotubes can have
semiconducting or metallic behavior, depending on the helical
arrangement of the carbon rings around the tubule \cite{saito}.
The metallic
nanotubes have all a typical band structure at low energies, in
which two different subbands cross at opposite Fermi points (in
the undoped system).

It is well-known that the so-called Luttinger liquid behavior
describes the regime with absence of electron quasiparticles
that is characteristic of one-dimensional (1D) electron systems
with dominant repulsive interactions \cite{emery,sol,hal}.
In the case of the undoped metallic
nanotubes, the fact of having four low-energy linear branches
at the Fermi level introduces a number of different scattering
channels, depending on the location of the electron modes near
the Fermi points \cite{berk}.
It has been shown, however, that processes which change the
chirality of the modes, as well as processes with large
momentum-transfer, are largely subdominant with respect to those
between currents of like chirality \cite{bal,eg}.
It has been concluded, therefore, that the metallic carbon
nanotubes should fall into the Luttinger liquid universality
class \cite{eg,kane}.
Supporting that expectation, characteristic experimental
signatures such as the power-law behavior of the tunneling
density of states have been measured in ropes \cite{bock} as well
as in individual single-walled nanotubes \cite{yao}.

%
%The unconventional electron behavior deviating from the Fermi
%liquid picture  \cite{unconv} and the properties of a
%Luttinger-liquid typical of a one-dimensional electron system
% \cite{sol,hal} have been also observed in carbon nanotubes
% \cite{bock}.

%In order to explain the behaviour of some two-dimensional (2D)
%systems (such as copper-oxide materials) a 2D Luttinger model was
%hypothesized  \cite{and}, but, at least for this model, the typical
%behaviour is lost as soon as one departs from $D = 1$
% \cite{ccm,arri} by an analytic continuation in the number $D$ of
%dimensions.

%Because the analysis of the Luttinger liquid behavior is usually
%made under the assumption of a local or short-range interaction
% \cite{sol,hal,vt,gz}, the unscreened Coulomb interaction (or some
%other singular interaction) was held responsible for the breakdown
%of the Fermi-liquid picture.

From the theoretical point of view, a relevant question is the
determination of the effects of the long-range Coulomb interaction
present in carbon nanotubes. In most part of the analyses
carried out for such systems, the electron-electron interaction
is taken actually as short-range, on the assumption that the
finite length of the experimental samples has to impose in
practice an infrared cutoff on the interaction. It is known,
however, that the Coulomb interaction is not screened in one
spatial dimension \cite{grab,bell},
and that it remains long-range even after
coupling a parallel array of 1D systems
---although the strength of the interaction can be consequently
reduced in this case \cite{egger,micro}.

The issue of considering the effects of the long-range Coulomb
interaction is significant, as they have been shown to lead
in general to unconventional electronic properties \cite{sing}.
Such effects have been shown to be responsible for a strong
attenuation of the quasiparticle weight in graphite \cite{ggv2}.
The long-range Coulomb potential $V(|{\bf r}|) \sim 1/|{\bf r}| $
yields the main electron interaction in carbon nanotubes as well
as in the two-dimensional (2D) layers of graphite, which have
a vanishing density of
states at the Fermi level. An important confirmation of the
marginal Fermi liquid behavior in such 2D layers comes from the
experimental measurements in graphite of a quasiparticle decay
rate linear in energy  \cite{exp}. Owing to the singular nature of
the Coulomb interaction, the imaginary part of the electron
self-energy in the 2D layer behaves like $g^2 \omega $
at weak $g$ coupling \cite{expl}. On the other hand, the effective
coupling scales at low energy like $g \sim 1/\log (\omega )$.
This fact explains that the quasiparticle weight is not
driven to zero by its logarithmic renormalization, being corrected
instead by terms of order $g^2 \log (\omega ) \sim 1/\log (\omega
)$ \cite{marg}.

The experiments on tunneling into the bulk of multi-walled
nanotubes (MWNTs)
have shown a power-law behavior of the tunneling
density of states, with an exponent $\alpha \sim 0.3$  \cite{B}.
This is close to the values measured in single-walled nanotubes
 \cite{bock,yao}, but it is much larger than expected in a wire
with a large number $N$ of subbands, where there should be a
significant reduction in the strength of the electron-electron
interaction. There have been attempts to confront this
puzzle under the assumption of a disorder-enhanced renormalization
of the interaction \cite{gogol}.
However, the exponents predicted in this treatment display a
typical $1/N$ reduction that make them to fall short with respect
to the experimental estimate \cite{B}. The enhanced
suppression of tunneling in the MWNTs supports
therefore the view that the long-range character of the Coulomb
interaction has to affect the system in a sizeable way.

In a recent Letter  \cite{0}, we developed an analytic
continuation in the number $D$ of dimensions to accomplish the
renormalization of the long-range Coulomb interaction at
$D \rightarrow 1$. This limit turns out to be highly singular,
since the marginal Fermi liquid behavior characteristic of
the 2D graphene sheet is established as soon as one departs from
$D = 1$. However, the attenuation of the electron quasiparticles
is increasingly strong as $D \rightarrow 1$, leading to an
effective power-law behavior of observables like the density of
states. In this way, we were able to predict a lower bound of
the corresponding exponent that turned out to be very close to the
value measured in experimental observations of the tunneling
conductance for MWNTs  \cite{B}.

In this paper, we focus on the singular behavior that arises at
the limit $D \rightarrow 1$ and introduce the effect of the
number of subbands that contribute to the low-energy properties
of carbon nanotubes. This issue is relevant for the investigation
of the nanotubes of large radius that are present in the MWNTs,
which are usually doped and may have a large number of subbands
crossing the Fermi level \cite{[21]}.
From a technical point of view, it becomes suitable the
implementation of a large-$N$ approach in the renormalization of
the electron system. Increasing the number of subbands has the
effect opposite to approaching the bare Coulomb interaction in
the limit $D \rightarrow 1$. We will see that it is possible to
take a kind of double scaling limit, $N \rightarrow \infty$
and $D \rightarrow 1$, in which the large-$N$ effects
are able to tame the singularities related to the long-range
interaction. Thus, regular expressions can be obtained for all
observables right at $D = 1$, bearing also a dependence on
the doping level of the nanotube system.

The property of the double scaling limit, of rendering finite
the effects of the renormalized long-range interaction, can be
explained by the onset of some sort of dynamical screening, in
the limit of an infinite number of subbands. Of course, in the
real carbon nanotubes of a multi-walled sample, the number of
subbands at the Fermi
level may be large but finite, and the singularity of the Coulomb
potential is not precisely reached due to the finite size of the
system. However, the double scaling limit describes the
universality class capturing the physics of the doped multi-walled
samples, as any other 1D approach to these systems is bound to
bear the divergence of the long-range interaction in the
low-energy limit.

In this paper we deal with a renormalization group (RG) approach,
in order to obtain the low-energy behavior of carbon nanotubes
with  long-range Coulomb interaction and a large number of
subbands at the Fermi level.
This approach is well-suited to the description of the scaling
properties of observables like the quasiparticle weight or
the density of states near the Fermi level. We will see that
the renormalization introduces a factor of $1/\sqrt{N}$ in
the critical exponent for those quantities.
Our results can be compared with those from the experiments
reported in Ref.  \cite{B}, where measurements of the tunneling
conductance have been carried out in doped MWNTs, with
a number of subbands at the Fermi level $N \approx 10$
(in the outer layer). For such values of $N$, and taking into
account the long-range Coulomb interaction, the critical exponents
that we obtain match well the values estimated from the experiments.

%%%%%%%%%%%%%%%%%%
\section{Band structure of graphene and carbon nanotubes}
%%%%%%%%%%%%%%%%%

We describe the band structure of carbon nanotubes by the
technique of projecting the band dispersion of a 2D graphite
layer into the 1D longitudinal dimension of the nanotube. The
2D band dispersion of graphene can be found in
Ref.  \cite{graph}. It consists of an upper and a lower branch that
only touch at the corners of the hexagonal Brillouin zone. Thus,
when the system is at half-filling, the metallic properties derive
from a pair of inequivalent Fermi points, around which there is
conical dispersion for the modes of the graphene sheet.

The 2D layers in graphite have a honeycomb structure
with a simple hexagonal Bravais lattice and two carbon atoms
in each primitive cell. We can start from the wavefunction
associated to the electronic density, where we introduce two
variational parameters, corresponding to two orbitals localized
around the two carbon atoms in the primitive cell. Then, by
introducing the wavefunctions in the energy functional for the lattice
hamiltonian, we can solve the variational problem and obtain the
following band dispersion, shown in Fig. 1(a) \cite{graph}:

\begin{equation}\label{gra}
  E({\bf k})=\pm \gamma \sqrt{1+4\cos^2(\frac{\sqrt{3}}{2}k_x)
  +4\cos(\frac{\sqrt{3}}{2}k_x)\cos(\frac{{3}}{2}k_y)}
\end{equation}

Since the basis of the honeycomb lattice contains two atoms,
there are two sublattices and two degenerate Bloch states at each Fermi point.
If we choose the Bloch functions separately on each
sublattice such that they vanish on the other, then  we can
expand the electron operator
in terms of the Bloch waves
\begin{equation}\label{expa0}
\Psi_\sigma(x,y) \sim \sum_{p\alpha}
 \exp( -i \alpha {\bf K \cdot r} )
\, F_{p\alpha\sigma} (x,y) \;
\end{equation}
where $\alpha=\pm$ labels the Fermi point,
  ${\bf r}=(x,y)$  lives
on the  sublattice $p=\pm$ under consideration and
$F_{p\alpha\sigma}(x,y)$ denote
slowly varying operators.
Thus, we can conclude that the low-energy excitations of the honeycomb
lattice at half-filling are described
by an effective theory of two 2D Dirac spinors \cite{graph}.

Starting from the graphene band structure (\ref{gra}), after
introducing periodic boundary conditions due to the cylindrical
geometry of the tube, we obtain the energy bands of a carbon
nanotube
\begin{equation}
E_{m} (k)= \pm \gamma \sqrt{1-4\cos(\frac{\pi
m}{N_b})\cos(\frac{\sqrt{3}k}{2}) +4{\cos^2(\frac{\sqrt{3}k}{2})}}
\label{peri}
\end{equation}
where $N_b$ is the number of periods of the hexagonal lattice
around the compact dimension (in the $y$ direction) of the cylinder.
So, at half-filling, metallic nanotubes have two Fermi points
corresponding to large momenta $\pm K_F$  (see Fig. 1(b)).
Now the low-energy expansion (\ref{expa0}) transforms
correspondingly and  the
electron operator is  written as  \cite{kane2}
\begin{equation}\label{expa}
\Psi_\sigma(x,y) = \sum_{p\alpha} (2\pi R)^{-1/2}
\exp( -i \alpha {\bf K \cdot r} )
\,\psi_{p\alpha\sigma} (x) \;
\end{equation}
which introduces
 1D fermion operators $\psi_{p\alpha\sigma}(x)$ depending
only on the longitudinal coordinate  $x$.

For later use, we note that
Eq. (\ref{peri}) implies a simple dependence of the number of
subbands that are found below the energy $E=\gamma\varepsilon $
$$
n_s(\varepsilon)
=\frac{N_b}{\pi}\arcsin(\sqrt{1-\varepsilon^2})\approx
\frac{N_b}{\pi} (\varepsilon+\frac{\varepsilon^3}{6}+...)
$$
At low energies, we can substitute the function $n_s(\varepsilon)$ with
the first term of its Taylor expansion without a large disagreement.

%%%%%%%%%%%%%%%%%%%%%%%%%%%%%%%%%%%%%%%%%%%%%%%%%%%%%%%%%%%%%%%%%%%%%%%%%%%%%%%
%%%%%%%%%%%%%%%%%%%%%%%%%%%%%%%%%%%%%%%%%%%%%%%%%%%%%%%%%%%%%%%%%%%%%%%%%%%%%%%
\section{Dimensional regularization near D = 1}
%%%%%%%%%%%%%%%%%%%%%%%%%%%%%%%%%%%%%%%%%%%%%%%%%%%%%%%%%%%%%%%%%%%%%%%%%%%%%%%
%%%%%%%%%%%%%%%%%%%%%%%%%%%%%%%%%%%%%%%%%%%%%%%%%%%%%%%%%%%%%%%%%%%%%%%%%%%%%%%
%%%%%%%%%%%%%
\setcounter{equation}0
In a previous paper  \cite{0}, we have developed an analytic
continuation in the number of dimensions in order to regularize
the infrared singularity of the long-range Coulomb interaction at
$D = 1$  \cite{ccm}. Our aim was to find the effective interaction
between the low-energy modes of metallic nanotubes, which have
linear branches crossing at the Fermi level. For this purpose, we
have dealt with the analytic continuation at general dimension $D$
of the linear dispersion around each Fermi point. We start then
with the hamiltonian
\begin{eqnarray}
H & = &  v_F  \sum_{\alpha\sigma}
    \int_0^{\Lambda } d p |{\bf p}|^{D-1}
     \int  \frac{d\Omega }{(2\pi )^D} \;
   \psi_{\alpha\sigma}^{+} ({\bf p}) \;
    \mbox{\boldmath $\sigma
  \cdot $} {\bf p} \;
            \psi_{\alpha\sigma} ({\bf p})   \nonumber   \\
   \lefteqn{  +   e^2 \int_0^{\Lambda } d p |{\bf p}|^{D-1}
     \int  \frac{d\Omega }{(2\pi )^D}  \;
   \rho ({\bf p})   \;   \frac{c(D)}{|{\bf p}|^{D-1}}  \;
          \rho (-{\bf p})  \;\;\;\;\;\; }
\label{ham}
\end{eqnarray}
where the $\sigma_i $ matrices are defined formally by $ \{
\sigma_i , \sigma_j \} = 2\delta_{ij}$. Here $\rho ({\bf p})$ are
density operators made of the electron modes
$\psi_{\alpha\sigma} ({\bf p})$, and
$ c(D)/|{\bf p}|^{D-1} $ corresponds to the Fourier transform of
the Coulomb potential in dimension $D$. Its usual logarithmic
dependence on $|{\bf p}|$ at $D = 1$ is obtained by taking the 1D
limit with
$c(D) = \Gamma ((D-1)/2)/(2\sqrt{\pi})^{3-D}$.

A self-consistent solution of the low-energy effective theory has
been found in  \cite{0} by determining the fixed-points of the
RG transformations implemented by the reduction
of the cutoff $\Lambda $. A phenomenological solution of the model
was firstly obtained \cite{pheno}, carrying a dependence on the
transverse scale needed to define the 1D logarithmic potential,
which led to scale-dependent critical exponents and prevented a
proper scaling behavior of the model \cite{pheno,wang}.
The dimensional regularization
approach of Ref.  \cite{0}, which we follow here, overcomes the
problem of introducing such an external parameter.

In the present paper, we incorporate the effect of having a
number $N$ of subbands crossing the Fermi level, which multiply
consequently the number of electron fields and terms in the
hamiltonian (\ref{ham}). Actually, as long as the long-range
Coulomb potential is strongly peaked at ${\bf p} = 0$, we will
focus on the dominant interactions that take place between currents
in which the electron modes remain in the same linear branch.

Thus, the one-loop polarizability $\Pi ({\bf k}, \omega_k)$ is
given by the sum of particle-hole contributions within each branch.
For a given subband, this leads in one spatial dimension to the
well-known result \cite{sol}
\begin{equation}
\left. \Pi ({\bf k}, \omega_k) \right|_{D = 1}
= \frac{2}{\pi } \frac{v_F {\bf k}^2}
 {  v_F^2 {\bf k}^2 - \omega_k^2  }\; \;
\end{equation}
For a number $N$ of different subbands, we take the analytic
continuation to general dimension $D$ \cite{iz}
\begin{equation}
\Pi ({\bf k}, \omega_k) = 2N b(D) \frac{v_F^{2-D} {\bf k}^2}
 { | v_F^2 {\bf k}^2 - \omega_k^2 |^{(3-D)/2} }\; \;
\label{pol}
\end{equation}
where $b(D) = \frac{2}{ \sqrt{\pi} } \frac{ \Gamma ( (D+1)/2 )^2
   \Gamma ( (3-D)/2 ) }{ (2\sqrt{\pi})^D \Gamma (D+1) }$.
In the case of carbon nanotubes we have $N = 2$, with the spin
degeneracy already taken into account in (\ref{pol}). In the
case of a thick nanotube with a large number of subbands,
the doping modifies the number of subbands to be taken into
account near the Fermi level.
Moreover, it is also conceivable that, in the
process of renormalization, $N$ may be given effectively by a
variable number of subbands depending on the energy cutoff, if
the relevant energy scale of the problem is large enough, as we
discuss in appendix.

In the large-$N$ limit we are interested in, the electron
self-energy $\Sigma ({\bf k}, \omega_k)$ is obtained by dressing
the bare Coulomb interaction with the polarization effects
given by (\ref{pol}). We have then, at the dominant level,
\begin{eqnarray}
\Sigma ({\bf k}, \omega_k)  & = &  - e^2 \int_0^{\Lambda }
     d p |{\bf p}|^{D-1}  \int \frac{d\Omega }{(2\pi )^D}
    \int \frac{d \omega_p}{2\pi }     \nonumber   \\
 \lefteqn{   G_0 ({\bf k} - {\bf p}, \omega_k - \omega_p)
 \frac{-i}{ \frac{|{\bf p}|^{D-1}}{c(D)} + e^2  \Pi ({\bf p},
    \omega_p) }     }
\label{selfe}
\end{eqnarray}
where $G_0$ is the bare electron propagator.

At general $D$, the self-energy (\ref{selfe}) shows a logarithmic
dependence on the cutoff at small frequency $\omega_k$ and small
momentum ${\bf k}$. This is the signature of the renormalization
of the electron field scale and the Fermi velocity. In the
low-energy theory, with high-energy modes integrated out, the
electron propagator becomes
\begin{eqnarray}
 \frac{1}{G}  & = &  \frac{1}{G_0} - \Sigma
     \approx    Z^{-1} ( \omega_k - v_F
  \mbox{\boldmath $\sigma
  \cdot $}
   {\bf k}) \nonumber \\
 &-&  Z^{-1} \frac{f(D)}{2 N}
 \sum_{n=0}^{\infty} (-1)^n g^{n+1}    \left(
   \frac{n(3-D)}{n(3-D)+2}  \omega_k    \right.\nonumber \\
  & + &  \left.  \left(1 - \frac{2}{D} \frac{n(3-D)+1}{n(3-D)+2}
   \right)   v_F
    \mbox{\boldmath $\sigma
  \cdot $}           {\bf  k}
      \right)
h_n (D)
   \log (\Lambda )
\label{prop}
\end{eqnarray}
where $g =2 N b(D) c(D) e^2 / v_F $, $f(D) = \frac{1}{ 2^D
\pi^{(D+1)/2} \Gamma (D/2) b(D) }$, $h_n (D) = \frac{ \Gamma
(n(3-D)/2 + 1/2) }
 { \Gamma (n(3-D)/2 + 1) }$.
The quantity $Z^{1/2}$ represents the scale of the bare electron
field compared to that of the renormalized electron field for
which $G$ is computed.

The renormalized propagator $G$ must be cutoff-independent, as it
leads to observable quantities in the quantum theory. This
condition is enforced by fixing the dependence of the effective
parameters $Z$ and $v_F$ on $\Lambda $ as more states are
integrated out from high-energy shells. We get the differential
RG equations for the effective parameters
$g$ and $Z$
\begin{eqnarray}
\Lambda \frac{d}{d \Lambda} \log Z (\Lambda )  & = &
 -  \frac{f(D)}{2 N} \sum_{n=0}^{\infty} (-1)^n g^{n+1}
     \;\;\;\;\;\;\;\;\;   \;\;\;\;\;\;\;\;\;    \nonumber  \\
  \lefteqn{  \frac{n(3-D)}{n(3-D)+2}   h_n (D) \equiv -\gamma (g)  }
\label{zflow}
\end{eqnarray}
\begin{eqnarray}
\Lambda \frac{d}{d \Lambda} g (\Lambda )  & = &
  \frac{f(D)}{2 N} \frac{2(D-1)}{D}g^2 \sum_{n=0}^{\infty} (-g)^n
   \;\;\;\;\;\;\;\;\;   \;\;\;\;\;\;\;\;\;
             \;\;\;\;\;\;\;\;\;     \nonumber  \\
  \lefteqn{ \left(  \frac{(3-D)n+1}{(3-D)n+2}
 \right)  h_n (D)  \equiv  -\beta(g)  }
\label{aflow}
\end{eqnarray}

The right-hand-side of Eq. (\ref{aflow}) is a monotonous
increasing function of $g $, for any dimension between 1 and 2.
At $D = 2$, the right-hand-sides of these equations can be summed
up, and they coincide for $N = 2$ with the functions obtained in
the case of the undoped graphene sheet \cite{marg}.
For $D = 1$ the function
$\beta (g)$ vanishes, so that the 1D model has formally a line
of fixed-points as it happens in the case of short-range
interaction. However, in the present model the variable $g$ is
sent to strong coupling in the limit $D \rightarrow 1$, and it
remains to be checked the behavior of the RG flow in that
regime.

%The number of subbands $N_S$ ($n$) appears in the RG equations,
%because of the polarizability $\Pi(k,\omega)$, which expressly
%depends on $N_S$.
%Since $\Pi$ depends on $b(D)$, we can introduce all
%dependence from $N_S$ in $b(D)$:
%$b(D,N_S)=b(D,N_0)\frac{N_S}{N_0}$.
%\frac{b(D,N_0)}{N_S}$$.
%Hence, $f(D)$ changes as follows: $f(D)\propto 1/b(D)
%\propto 1/N_S$.
%Thus, we can summarize the rescaling due to the number of subbands after
%introducing $$ b\rightarrow n_s b_I \qquad f\rightarrow
%\frac{f_I}{n_s} \qquad \beta\rightarrow \frac{\beta_I}{n_s} \qquad
%\tau\rightarrow \frac{\tau_I}{n_s} \qquad g\rightarrow n_s
%g_I\qquad$$ where we label with $I$ the value when $n_s=4$.
%
%The rescaling of $g$ is for the initial value $g_0$, while the
%renormalized one $g(\Lambda)$ from the RG equations has the
%scaling factor in the $\beta$ function.

In order to approach the limit $D \rightarrow 1$, we have to look
for the asymptotic dependence on $D$ of the functions appearing in
the RG equations. We will see that this dependence appears as
$D-1$ and $D-3$ factors, revealing that these are the two critical
dimensions, corresponding to a marginal and a renormalizable
theory, respectively.

Starting with the function $\beta (g)$, we need to carry out the
sum at the right-hand-side of Eq. (\ref{aflow}). This is given in
terms of the hypergeometric special functions
\begin{equation}
B_D (g) = \sum_{n=0}^{\infty} (-g)^n
   \left(  \frac{(3-D)n+1}{(3-D)n+2}  \right)  h_n (D)
\end{equation}
Since we need the $\beta $ function near $D = 1$, we can
approximate $B_D (g)$ with the simple function $B_1 (g)$
\begin{equation}
B_1 (g) = \frac{\sqrt{\pi }}{g}
    \left( 1 - \frac{1}{\sqrt{1+g}} \right)
\end{equation}
To first order in $D - 1$, we get then
\begin{equation}
\beta (g) \approx - \frac{f(D)}{2 N} \frac{2(D-1)}{D}
        \sqrt{\pi }g \left( 1 - \frac{1}{\sqrt{1+g}} \right)
\label{beta1}
\end{equation}

As long as the coupling $g$ depends implicitly on the
dimension $D$, we also need to know its asymptotic behavior
near $D = 1$. In practice, we can translate the analysis to the
initial value of the coupling, $g_0 (D)= 2 N b(D) c(D) e^2/v_F$.
Carrying out an expansion near $D=1$, we obtain
\begin{equation}
\frac{1}{g_0(D)}\approx  \frac { 4 \pi^{2}v_F}{2 N e^2}
(D-1)[\frac{D}{2}-\frac{\zeta(3)}{4}(D-1)^3]
\end{equation}
where $\zeta(s)=\sum_{k=1}^\infty k^{-s}$ is the Riemann zeta
function.
So we can use in our calculations the simple expression
\begin{equation}
  g_0(D)\approx 2 N \frac{ e^2} { 4 \pi^{2}v_F} \frac{2}{(D-1)D}
\label{eqg0}
\end{equation}

Finally, it can be seen that the right-hand-side of
Eq. (\ref{zflow}) does not behave as a
simple power law. This is due to the fact that the terms of the
series vanish too slowly when $n$ increases. The $\gamma $
function can be expressed in the form
\begin{equation}
\gamma (g) =   \frac{f(D)}{2 N}
\frac{3-D}{2}g T_D(g)
\end{equation}
where the series $T_D(g)$ is available on tables. For $D=1$
it has the simple expression
\begin{equation}
T_1(g)=\sqrt{\pi}\frac{2\sqrt{g+1}-g-2}{g\sqrt{g+1}}
\end{equation}
The scaling of the electron wave function near $D = 1$ is
therefore given by
\begin{equation}
\gamma (g) \approx   \frac{f(D)}{2 N}
 \sqrt{\pi }
 \left( 2 - \sqrt{1+g} - \frac{1}{\sqrt{1+g}}  \right)
\end{equation}
This coincides formally with the anomalous dimension that is found
at $D = 1$ in the exact solution of the Luttinger model,
what provides an independent check of our RG approach to the
1D system.

\section{RG scaling and low-energy density of states}
\setcounter{equation}0

%Now we can rewrite the second RG equation (\ref{zflow})
%as follows:
%\begin{equation}
%\frac{d \log(Z)}{dx}= \frac{f(D)}{2 N}
%\frac{3-D}{2} T_D(g(x))g(x)
%\end{equation}

Now we can rewrite the RG Eq. (\ref{aflow}) near $D = 1$ as
follows:
\begin{equation}
\frac{d g}{dx} = - \frac{f(D)}{2 N}
\frac{2(D-1)}{D}
        \sqrt{\pi }g \left( 1 - \frac{1}{\sqrt{1+g}} \right)
\label{flow1}
\end{equation}
where $x=-\log(\Lambda)$. At $D = 1$ we have formally from
Eq. (\ref{flow1}) a line of fixed-points, which cover all
values of $e^2/v_F$. In spite of the fact that $g$ goes to strong
coupling as $D \rightarrow 1$, we can see that the behavior in
Eq. (\ref{eqg0}) does not suffice to cancel the $D-1$ factor at
the right-hand-side of Eq. (\ref{flow1}). Therefore, we make sure
in this way that the long-range Coulomb interaction is at a
fixed-point of the RG at $D = 1$.

We can also rewrite the RG equation  (\ref{zflow})
as follows:
\begin{equation}
\frac{d \log(Z)}{dx} = \frac{f(D)}{2 N}
 \frac{3 - D}{2}  T_D(g) g
\label{ag}
\end{equation}
Near $D = 1$ we find a crossover  to a
behavior with a sharp reduction of the quasiparticle weight
in the low-energy limit $\Lambda \rightarrow 0$. All
that is displayed in Fig. 5 in appendix, where we have represented the
electron
field scale square $Z$. If $D$ is above $\approx 1.2$, we have
a clear signature of quasiparticles in the nonzero value of $Z$ at
low energies, whereas for lower values of $D$ the picture cannot be
distinguished from that of a vanishing quasiparticle weight.

Our target is to compare theoretical results with measurements
of the tunneling density of states carried out in nanotubes of
large radius \cite{B}. The density of states $n(\varepsilon )$
computed at dimensions between $1$ and $2$ displays an
effective power-law behaviour which is given by
\begin{eqnarray}
n( \varepsilon ) &\approx & 2 N Z( \varepsilon ) |\varepsilon |^{D-1}
\Rightarrow  \nonumber \\
\log(n( \varepsilon ) ) &\approx &
\log(2 N) + \log{Z( \varepsilon )}+
(D-1) \log(|\varepsilon |) \label{l8}
\end{eqnarray}
The last term in Eq. (\ref{l8}) comes from the analytic
continuation of the conical dispersion, which gives rise at
$D = 2$ to the well-known behavior of the graphite
layer, $n( \varepsilon ) \sim |\varepsilon |$.

In order to obtain the linear dependence of
$\log(n( \varepsilon ))$ on $\log(|\varepsilon |)$,
we have to introduce the low-energy behavior of
$Z( \varepsilon )$ in Eq. (\ref{l8}), by which
\begin{equation}\label{aster}
  \log(n( \varepsilon ) )\sim  \left( \alpha_Z - (D-1)
\right) x \equiv \alpha_D x
\end{equation}
If we want to keep the linear behaviour of $\log(Z)$ in
$x=-\log(|\varepsilon |)$, we have to make an approximation. We
can solve the RG equation (\ref{zflow}) near
$x = 0$ with a simple first
order expansion, $ \log(Z) \sim \gamma (g_0 ) x$, where
$g_0$ is the initial value of the coupling. Then we can
analyze the dependence of $\alpha_Z$ on the dimension $D$ and
at large $N$ starting from the expression
\begin{equation}\label{alphaz1}
  \alpha_Z \approx \frac{f(D)}{2 N} \frac{3-D}{2} T_D(g_0)g_0
\end{equation}

\subsection{Analytic continuation near D=1}

Following Ref. \cite{0}, we can obtain a lower bound for the
exponent of the density of states, for the purpose of
comparison with
experiments, by estimating the minimum of the absolute
value of $\alpha_D$, for dimensions ranging
between $D=1$ and $D=2$. The evaluation can be carried out using
Eq. (\ref{aster}),
after plugging the expression of $\alpha_Z$ in (\ref{alphaz1}),
with its strong dependence on $D$.

In order to obtain an analytical expression for $\alpha_D $, we
can approximate  $T_D(g)$,  starting from  $T_1(g)$, by using
for $1<D<2$ the simple approximation
$T_D(g)\approx \frac{D+1}{2}T_1(g)$. %(in Fig. we show the validity
%of this approximation for $g=1$).%
Thus, we can give a simple formula for $\alpha_D $ rewriting Eq.
(\ref{aster}) in the form
\begin{equation}\label{alpha}
  \alpha_D \approx g_0T_1(g_0)\frac{(3-D)f(D)(D+1)}{8 N}-(D-1)
\end{equation}

We obtain a minimum value for $|\alpha_D |$ as a function of $D$
when we introduce the
expression of $g_0(D)$ obtained in the previous section.
From Fig. 2 we can see that the maximum value for
$\alpha_D$ ($\alpha_M$) corresponds to a dimension between $1$ and
$1.2$. If the number of subbands is increased, then the value of
$|\alpha_M |$ decreases while the corresponding dimension approaches
$1$.

Our analytic computation allows to understand the interplay
between the number of subbands $N$ and the dimension $D$ at which
the minimum of the critical exponent $|\alpha_D |$ is obtained.
We find that this is around $|\alpha_D | \approx 0.3$ for not too large
values of $N$, which is in agreement with the values
measured experimentally. However, the fact that the minima are
obtained above $D = 1$ stresses once again the singular character
of the 1D Coulomb interaction. The physical description can be
clarified by noticing that the purely 1D limit can be approached
by increasing the number $N$ of subbands, as we propose below.

%Such a result
%demonstrates that the picture of a thick nanotube as an
%aggregate of 1D channels does not allow to obtain the correct
%values of the critical exponents. So we need an appropriate
%description of the dimensional crossover between one and two
%dimensions.

\subsection{Double scaling approach}

Here we exploit the dependence on the number of subbands $N$ as
a way of regularizing the divergent effects of the Coulomb
interaction, when $D$ approaches the critical dimension $D=1$.
As it is easily seen inspecting the $\alpha_Z$ function in Eq.
(\ref{alphaz1}), a large $N$ can be combined with the divergence
in $D - 1$ from the coupling constant $g_0$ in Eq. (\ref{eqg0}),
so that both quantities regularize each other in a sort of double
scaling limit. From the physical standpoint, this mechanism of
regularization can be regarded as a kind of screening of the
singular Coulomb interaction at $D = 1$.

We use the above method of regularization near the critical
dimension $D=1$, in order to give an alternative evaluation of
the critical exponent $\alpha $. In the limit of large $N$,
the dimension where the crossover effectively takes place,
corresponding to the maximum value of $\alpha_D $, approaches
$D=1$. In this regime, the last term in Eq. (\ref{alpha})
vanishes and we can use the value of $\alpha_Z$ as an estimate
of $\alpha$.

In the limit of large  $N$ and small values of $D - 1$,
we have $T_1(g) \rightarrow -\sqrt{\frac{\pi}{g}}$, so that
Eq. (\ref{alphaz1}) becomes
\begin{equation}
  \alpha_Z\approx -
   \frac{f(D)(3-D)(D+1)}{8 N} \sqrt{\pi \: g_0}
\label{alphazl}
\end{equation}
Taking into account the form of $g_0$ in Eq. (\ref{eqg0}),
we see that $\alpha_Z$ vanishes at large $N$ as $\sim 1/\sqrt{N}$,
and it diverges at $D \rightarrow 1$ as $\sim 1/\sqrt{D - 1}$.

Now we can use the regularization from the number of
subbands, in order to access a physical regime that would
be otherwise hidden in the usual treatment of the 1D
Coulomb interaction. In other words, we take simultaneously
the limits $N \rightarrow \infty $ and $D \rightarrow 1$ by
keeping a finite
renormalized value $N_{ren}$ for the number of subbands, $N_{ren}
\equiv N (D-1)$. We use then $N_{ren}$ as a way of parameterizing
the dependence of the critical exponent $\alpha$ on the doping
level, which sets the number of subbands at the Fermi level in the
carbon nanotubes. From Eq. (\ref{alphazl}) we obtain a simple
expression for $\alpha$ as a function of the renormalized number of
subbands
\begin{equation}
  \alpha = -\frac{1}{4}\sqrt{\frac{ e^2 }{ \pi^{2}v_F N_{ren} }}
\label{nren}
\end{equation}

The agreement between the two estimates of the critical
exponent $\alpha $ by the double scaling limit and by analytic
continuation near $D = 1$ is shown in Fig. 3(a).
The dependence of $\alpha_Z$ on the number of subbands in the
second approach can be actually borne out by numerical results.
These are in good agreement with our analytical
predictions, as we show in Fig. 4. The dependence of $\alpha_Z$ on
the number of subbands is rather similar to $1/\sqrt{N}$,
especially for dimensions near $1$.
Thus, from this overall picture we can predict that, by changing
the doping level in carbon nanotubes of large radius, a sizeable
variation of the anomalous electron dimension should be observed
in the range $N \sim 2 - 100$,
with a consequent change in the power-law behavior of the
tunneling density of states given by Eq. (\ref{nren}).

\section{Discussion}

In this paper we have described the appropriate framework to
deal with the low-energy effects of the long-range Coulomb
interaction in 1D electron systems. Our interest has been
focused in the scaling behavior of quantities like the
quasiparticle weight or the low-energy density of states,
which can be compared directly with the results of transport
experiments. For this purpose, we have developed a RG approach
in a dimensionally regularized theory devised to interpolate
between the 2D conical band dispersion of graphene and the 1D
band dispersion of the carbon nanotubes. The analytic
continuation in the number of dimensions has allowed to avoid
the infrared singularities that the long-range Coulomb interaction
poses at $D = 1$, providing insight at the same time about the
fixed-points and universality classes of the theory in the limit
$D \rightarrow 1$.

Thus, we have seen that the 1D electron system with long-range
Coulomb interaction is at a fixed-point of the RG flow, for
arbitrary values of the interaction strength. However, the
scaling behavior of observables like the quasiparticle weight
is dictated by divergent critical exponents in the limit
$D \rightarrow 1$. In order to make contact with the results
of transport experiments on MWNTs,
we have introduced the relevant dependence on the number $N$
of subbands at the Fermi level. By increasing the value of $N$,
it has been possible to approach the critical dimension $D = 1$
while producing finite, sensible estimates of the critical
exponents in the carbon nanotubes. The double scaling limit,
$N \rightarrow \infty $ and $D \rightarrow 1$, gives rise to
a phase of the system that captures the physical properties
measured in nanotubes of large radius, which use to be
significantly doped. In this way, we have been able to describe
a new universality class that arises in the large-$N$ limit,
in which the effect of the large number of subbands is to
regularize the singular 1D Coulomb interaction. The new phase
arises from a sort of 1D screening of the long-range interaction,
giving rise to a modified Luttinger liquid picture in which the
critical exponents depend on the doping level as well as on the
strength of the interaction.

The framework that we have introduced is appropriate for the
description of carbon nanotubes of large radius, in which the
shift of the Fermi level leads to a sizeable variation of the
number of subbands contributing to the low-energy electronic
properties. Such a situation is realized in the MWNTs, which are
concentric arrangements of carbon nanotubes with a diameter of
$\sim 10 \; {\rm nm}$ and typically $\sim 5$ inner cylinders.
The transport measurements carried out in the MWNTs reflect
usually the electronic properties of the outer layer, to
which the electrodes are attached. In these systems the number
of subbands at the Fermi level can be large, i.e. of the order
of $\sim 10-20$ taking into account the electron spin, because of
the electrochemical doping.

One of the most significant
observations made in the MWNTs has been the power-law behavior
of the tunneling conductance as a function of the temperature
or the bias voltage. In the experimental measurements, the
maximum energy scale reached in those variables has been
always below $100 \; {\rm meV}$. This is of the order of
the typical energy spacing between the different subbands,
so that the transport
measurements  refer to the contribution of a fixed number of
subbands at low energies. In the case that the experiments were
carried out over a larger energy range, the physical
description should be modified to allow for a variable number of
subbands depending on the cutoff energy, producing the results
that are shown in appendix.

The measurements carried out in the MWNTs have displayed a
power-law behavior of the tunneling conductance, that gives
a measure of the low-energy density of states, with exponents
ranging from 0.24 to 0.37 \cite{B}. These values are, on the
average, below those measured in the single-walled nanotubes,
which are typically about $\approx 0.35$  \cite{yao}.
We see that our results can
account satisfactorily for this slight reduction in the critical
exponent with the change of the nanotube thickness. We observe
from Fig. 3(b) that, for a number of subbands $N = 2$, the value
of $|\alpha |$ corresponding to an undoped single-walled
nanotube is slightly below 0.3. This means that our choice for
the interaction strength (with some reduction from the dielectric
constant) has been appropriate. The important point is that, by
varying the number $N$ of subbands from 2 to 10, we obtain a shift
in the value of the critical exponent which is in fair agreement,
on the average, with that observed from the different transport
experiments. This lends strong support to our picture of the
renormalization of the long-range Coulomb interaction.

The main prediction that comes from our study is that there should
be a significant reduction in the critical exponent of the
tunneling density of states as the doping level is increased in
nanotubes of large radius. We have seen that the presence of a large
number of subbands at the Fermi level implies a reduction of $|\alpha|$,
by a factor of the order of $N^{-1/2}$. It would be of considerable
importance to test such a dependence in experiments carried out using
samples with different amount of doping. That kind of effect should
arise also as a consequence of considering wider compact nanotube
structures, stressing in another way the significance of the geometry
in the electronic properties of the nanotubes.

% To summarize, we have studied MWNTs,
%i.e. carbon tubes with $10$ $nm$ diameter which have typically
%$\sim 5$ inner cylinders. In similar systems the number of
%occupied subbands can be large, i.e. of order $N\sim10-20$,
%instead of $4$, because of the electrochemical doping  \cite{[21]}.
%The effects of this different number of subbands is expected to
%decrease the value of the exponent for MWNTs substantially, in
%comparison with the case of SWNTs. In the experiments the
%suppression of $\alpha$ is different from the expectations
% \cite{B}: the exponents for bulk and end tunneling of MWNT are
%often close to SWNT values ranging between $0.26$ and $0.38$ for
%bulk exponent and between $0.5$ and $1.1$ for end  exponent.

%The model for a doped nanotube requires the introduction of a large
%number of subbands. The latter corresponds to a sort of regularization of
%the divergences appearing at the critical dimension $D=1$ because of
%the Coulomb interaction. In the limit of an infinite number of
%subbands, the crossover dimension approaches $D=1$.
%Both analytical and numerical calculations show that
%the effect of the number of subbands amounts to a reduction of the bulk
%exponent. Our earlier calculations predicted $\alpha\sim 0.26$ for the
%usual
%4-subbands system  \cite{0}, while the introduction of a large
%number of subbands implies a reduction of $\alpha$, according to a
%power law $N_s^{-1/2}$. It would be of considerable importance
%to test such a dependence in experiments carried out using samples with a
%various amount of doping.

\vspace{0.5cm}

\noindent{\bf Acknowledgement}\\

\noindent This work was partly supported by the Programma di Ricerca di
Interesse Nazionale cofinanced by MIUR (Es. Fin. 2002) "Effetti di spin,
interazione e proprieta` di trasporto in sistemi elettronici fortemente
interagenti a bassa dimensionalita`".

\newpage
\appendix

\section{Number of subbands depending on cut-off energy}
\setcounter{equation}0

Next, we consider a number of subbands $N_S$ depending on the cut-off.
So, we introduce the
energy dependent function $n_s(\varepsilon)$ which substitutes
the usual $n_s$. We define a unit energy as $\varepsilon_p$, so
that we have $x_p=-\log(\varepsilon_p)=0$. The corresponding
number of subbands
reads $N_s(\varepsilon_p)=N_0+N_p$  \cite{notaa}
\begin{eqnarray}
N_s(\varepsilon)&=&N_0+N_p \frac{\varepsilon}{\varepsilon_p}
\Rightarrow \nonumber \\
N_s(x)&=&N_0+N_p e^{-x} \Rightarrow n_s(x)=1+n_p e^{-x}\nonumber
\end{eqnarray}
Here $n_p$  represents the number of subbands, in a unit of energy
normalized with respect to $N_0$.

When $n_s$ depends on the energy we obtain, by increasing the number of
subbands, that $\alpha_Z$ vanishes as $n_s^{-1/2}$. However a stronger
$\alpha_N$ contribution causes a growth of $\alpha$ (although its
absolute value stays less than $1$). The relevant correction in
the eq. (\ref{aster}) is due to the direct contribution of the
number of subbands depending on the energy. This is the $\alpha_N$
term which can strongly modify the $\alpha_D$ value. Near $x=0$, we
can obtain this term as a simple function of $n_p$, which
represents the number of subbands, in a unit of energy normalized
with respect to $N_0$
$$
\alpha_N \approx -\frac{n_p}{1+n_p}
$$
This term does not depend on the dimension, so it does not affect
the search of the maximum in the $\alpha_D$, rather it gives a simple
shift in the global $\alpha$ value, which we have to compare with
the experimental value of the parameter. Larger values of $n_p$ give larger
contributions (in absolute value) from this term which ranges between $0$
and $1$ (once again, in absolute value).

In the case of a linear dependence of the number of subbands on the
energy, a large correction to the measurable $\alpha$ exponent
can arise from the $\alpha_N$ contribution ($\alpha_N\sim 0-1$),
while the effective parameter $Z$ yields strong deviations from
the behaviour at $n=4$, even when the crossover dimension
approaches the fixed point $D=1$.

Some interesting effects
regarding the numerical calculation of $Z$ arise, when the number of
subbands depends on the cutoff energy ($n_p$ is the density of the
number of subbands in the energy unit). In this case a strong
deviation from the usual behaviour appears. The renormalization of
the electron field scale $Z$, for a fixed number of subbands (4
subbands, i.e. $N=2$),
shows a drastic suppression of the electron field scale, which
takes place over a variation of only two orders of magnitude in
the energy scale. When the number of subbands increases, we
observe that the scale of energies, and also the shape of the curve,
are quite different. The energy scale goes towards lower and lower
energies, while a large plateau appears in the curve
$Z(-\log(E))$. A simple mechanism can explain this behaviour. When
the cutoff increases, many subbands have to be taken into account
and the system, at any dimension greater than $1$, is more similar
to a 2-dimensional graphene sheet than to a 1-dimensional wire.
Only when the energy goes below a small value (i.e. $x>x_T$) the
system restores its Luttinger liquid behaviour and the
quasiparticle amplitude vanishes, following the usual behaviour.
This is what we can observe on the left of Fig. 5 . All that is clear also
on
the right, where we show the dependence of $\alpha_Z \approx \frac{d
\log(Z)}{dx}$ on $x$. Here the asymptotic behaviour reflects the
small $n_p$ results, whereas the values for `high energies` are
quite similar to those of the usual 2D graphene.

All the phenomenology of a tube with a very large number of
subbands below the cutoff energy is more similar to that of 2D
graphene than to a 1D Luttinger liquid. This effect
could be seen either as a smaller crossover dimension, or as the
persistence of quasiparticle poles at relatively high energy. We
can obtain the same physics if we imagine the radius of the tube
going to infinity: the geometrical limit of the tube is the
graphene layer, while the energy spacing between two subbands
vanishes, so that an infinite number of subbands have to be taken
into account, also for very small energies.

\newpage
%%%%%%%%%%%%%%%%%%%%%%%%%%%%%%%%%%%%%%%%%%%%%%%%

\section*{Figure Captions}

{\bf Fig. 1}: (a) Energy levels in a graphite sheet: the cusps appear at the
six corners of the first Brillouin zone.
(b) Low-energy band structure of metallic   carbon nanotubes with different
radius.
The energy is measured in
units of the hopping parameter and the momentum in units of the
inverse lattice spacing.
 As shown in (c), the radius is connected to the value of $N_b$ in a simple
way: $R=\sqrt{3}a/(2 \pi \sin(\pi/N_b))\approx N_b\sqrt{3}a/(2 \pi)$,
 where  $a$ denotes
the honeycomb lattice constant ($a/\sqrt{3}=1.42${\AA}).\\
%%%%%%%%%%%%%%%%%%%%%%%%%%%%%%%%%%%%%%%%%%%%%%
%%%%%%%%%%%%%%%%%%%%%%%%%%%%%%%%
\\
\noindent
{\bf Fig. 2}: In (a) we show $\alpha$ as a function of the
dimension (Eq. (\ref{alpha})), and we choose the maximum value of
this function between $D=1$ and
$D=2$ as an estimate of the critical exponent.
We find a value
for $\alpha$ in the usual $N=2$ undoped nanotube that reproduces
the anomalous exponent measured
experimentally ($\alpha\approx - 0.3$), corresponding to a
dimension for the crossover
between $1.1$ and $1.2$.
The value of $\alpha$ is compared to the term $\alpha_Z(D)$
(dashed line) dominant near the divergence $D\rightarrow1$.
As shown in (b), when the number of subbands increases, the
absolute value of the maximum decreases,
while the corresponding dimension approaches $1$
(each curve corresponds to a different value of $N$).\\
%%%%%%%%%%%%%%%%%%%%%%%%%%%%%%%%
%%%%%%%%%%%%%%%%%%%%%%%%%%%%%%%
\\
\noindent
{\bf Fig. 3}: In this paper we propose also a different way to estimate the
critical exponent, as the value of $\alpha_Z$ in the double limit
$N \rightarrow \infty $ and $D \rightarrow 1$.
A comparison between the results of the double scaling approach
(dashed line) and those from the analytic continuation in $D$
(full line) is shown in (a). The increase of $\alpha$
for a moderate number of
subbands is shown in (b). We have taken in all cases a value of
the coupling $e^2 /(\pi^2 v_F) \approx 0.5$, which is appropriate
for MWNTs, owing to the reduction in this value due to the
interaction with the inner metallic cylinders  \cite{egger}.
We can conclude that, when $N$ is very large, there is a good
agreement between the results obtained with the two different
approaches, i.e. double scaling and analytic continuation.
The dashed portion of the curve in (b) agrees, in the one-mode limit
(i.e. for $N=0.5$), with the results of Ref. \cite{0}.\\
%%%%%%%%%%%%%%%%%%%%%%%%%%%%%%%
%%%%%%%%%%%%%%%%%%%%%%%%%%%%%%%%%%%%%%%%%%%%%%%%
\\
\noindent
{\bf Fig. 4}: On the left, we show the linear dependence
of $\alpha_Z$ on the inverse square root of $N$, for different
values of $D$. On the right, we observe a rather smooth behavior
of the slope of the curves for each dimension in the first
figure.\\
%%%%%%%%%%%%%%%%%%%%%%%%%%%%%%%%%%%%%%%%%%%%%%%%
%%%%%%%%%%%%%%%%%%%%%%%%%%%%%%%%%%%%%%%%%%%%%%%%
\\
\noindent
{\bf Fig. 5}: On the left, we plot $Z(-\log(E))$ for a different number of subbands,
depending on the energy. On
the right, we show the dependence of $\alpha_Z \approx \frac{d
\log(Z)}{dx}$ on $x$.\\
%%%%%%%%%%%%%%%%%%%%%%%%%%%%%%%%%%%%%%%%%%%%%%%%

\begin{thebibliography}{99}
\bibitem{notaa}
We use a simple formula we obtained from the Taylor
expansion of the $n_s(\varepsilon)$ function in section 2. However, we
have to implement a correction, because we know that, when $E$
is set to $0$, we have $N_s(0)=N_0=4$.
\end{thebibliography}
\end{document}